\documentclass[aps,12pt,final,notitlepage,oneside,onecolumn,nobibnotes,nofootinbib,
superscriptaddress,noshowpacs,centertags]{revtex4}

\usepackage{graphicx}
\usepackage{amssymb}
\usepackage{bm}

\newcommand{\be}{\begin{equation}}
\newcommand{\ee}{\end{equation}}
\newcommand{\bel}[1]{\be\label{#1}}
\newcommand{\re}[1]{Eq.~(\ref{#1})}
\newcommand{\ds}{\displaystyle}
\newcommand{\hsp}{\hspace*{1pt}}
\newcommand{\ela}{E_{\rm\hsp lab}}

\def\refitem#1{\relax}

\begin{document}
\title{Hydrodynamic modeling of deconfinement phase transition in nuclear collisions}

\author{\firstname{I.N.} \surname{Mishustin}}
\email{mishustin@mbslab.kiae.ru}
\affiliation{Frankfurt Institute for Advanced Studies,~D--60438 Frankfurt am Main, Germany}
\affiliation{The Kurchatov Institute, Russian Research Center, 123182 Moscow, Russia}

\author{\firstname{A.V.} \surname{Merdeev}}
%\email{merdeev@mbslab.kiae.ru}
\affiliation{The Kurchatov Institute, Russian Research Center, 123182 Moscow, Russia}

\author{\firstname{L.M.} \surname{Satarov}}
%\email{satarov@mbslab.kiae.ru}
\affiliation{Frankfurt Institute for Advanced Studies,~D--60438 Frankfurt am Main, Germany}
\affiliation{The Kurchatov Institute, Russian Research Center, 123182 Moscow, Russia}

\begin{abstract}
    The (3+1)--dimensional ideal hydrodynamics is used to simulate collisions
    of gold nuclei with bombarding energies from 1 to 160 GeV per nucleon.
    The initial state is represented by two cold Lorentz-boosted nuclei.
    Two equations of state:
    with and without the deconfinement phase transition are used.
    We have investigated dynamical trajectories of compressed baryon-rich
    matter as functions of various thermodynamical variables.
    The parameters of collective flow and hadronic spectra are calculated.
    It is shown that presence of the deconfinement phase transition leads to
    increase of the elliptic flow and to flattening of proton rapidity
    distributions.
\end{abstract}

\maketitle

\section{Introduction}
    Relativistic hydrodynamics is a very popular approach to describe high--energy nuclear collisions.
    Especially useful is its sensitivity to the equation of state (EoS)
    of strongly interacting matter and, in particular, to its phase transitions.
	In fact, extracting this EoS is
    the main goal of heavy--ion experiments.
    Some signatures of a deconfined quark-gluon plasma (QGP)
    have been already found in RHIC experiments with c.m. bombarding
    energies \mbox{$\sqrt{s_{NN}}=60-200$ GeV}.
    Presumably, this plasma can be also created at lower (SPS, AGS) energies.
	A more detailed data should be
    obtained in the low--energy runs at RHIC and future FAIR and NICA experiments.

    A large amount of experimental data on nuclear collisions at AGS, SPS and RHIC energies
    have been successfully described by different versions of the hydrodynamic model.
    The first model of this kind has been proposed by Landau~\cite{Lan53}.
    One can roughly divide the existing versions of
    ideal hydrodynamics into two classes. The first class includes the models which apply fluid
    dynamical simulations from the very beginning, i.e. starting from cold equilibrium nuclei.
    The models of the second class introduce an~excited
    and compressed initial state -- a locally equilibrated ''fireball'' formed at an early
	non-equilibrium stage of the collision. Up to now many versions of
    the fireball--based hydrodynamic model were developed ranging from simplified
    (1+1)-- and
    \mbox{(2+1)--dimensional} models to more sophisticated
    (3+1)--dimensional ones (see ~\cite{Sat07} and references therein).

    Historically, early 3D hydrodynamic models of relativistic nuclear collisions~\cite{Ams75,Sto79}
    used cold Lorentz-contracted nuclei in the initial state.
	However, such models become less justified at high bombarding energy
	due to the nuclear transparency effects.

    One should have in mind that the hydrodynamical approach can not be directly applied to late
    stages of a heavy--ion reaction when binary collisions of particles become too rare to maintain
    the local thermodynamic equilibrium. A usual way~\cite{Coo74} to circumvent this difficulty
	to introduce a so called ''freeze--out'' criterium to stop the hydrodynamic description at late times.
    A more consistent procedure has been
    proposed in Refs.~\mbox{\cite{Bas00,Pet08}} within a hybrid ''hydro--cascade'' model.

    In this paper we formulate a version of the ideal (3+1)--dimensional hydrodynamics suitable
    for the domain of NICA--FAIR energies. This model belongs to the first class
    and uses a new EoS~\cite{Sat09} with the deconfinement and liquid--gas phase transitions.
    By performing simulations with and without the deconfinement phase transition we try to find
	the observables
    which are sensitive to creation of the QGP.
    A more detailed description of the results will be given in Ref.~\cite{Mer10}.

\section{Formulation of the model}
    Below we study the evolution of highly excited, and possibly deconfined, strongly--interacting
    matter produced in ultrarelativistic heavy-ion collisions. It is assumed that this evolution can
    be described by the equations of ideal relativistic hydrodynamics~\cite{Lan87}:
\begin{eqnarray}
            &&\partial_\nu T^{\mu\nu}=0\hsp,
        \label{emcon}\\
            &&\partial_\mu (nu^\mu)=0\hsp.
        \label{chcon}
\end{eqnarray}
    Here~$T^{\mu\nu}$ is energy-momentum tensor ($\hbar=c=1$)
        \bel{emten}
            T^{\mu\nu}=(\varepsilon+P)\hsp u^\mu u^\nu -P\hsp g^{\hsp\mu\nu},
        \ee
    $\varepsilon, n$ and $P$ are the rest--frame energy density, the net baryon density and
    pressure of the fluid,
    $u^\mu = (\gamma, \gamma\hsp\bm{v})^{\hsp\mu}$ is its collective 4--velocity,
	$\gamma = (1 - \bm{v}^2)^{-1/2}$,
    $\bm{v}$ is the fluid 3--velocity, $g^{\hsp\mu\nu}$ is the metric tensor.
    To solve Eqs. (1)--(3) it is necessary to specify the
    EoS $P=P(n,\varepsilon)$ of the fluid and initial conditions.

    In our calculations we use the EoS of strongly interacting matter
    with the first order deconfinement phase transition (EoS--PT)~\cite{Sat09}.
    The hadronic phase is described as the hadron resonance gas
    including known hadrons with masses up to 2 GeV.
    Finite size effects are taken into account by introducing the excluded volume
    corrections. The same excluded volume parameter \mbox{$v_e=1$ fm$^3$} is used
    for all hadronic species. A Skyrme--like mean field \mbox{$U=U(n)$}
    is added to guarantee that the hadronic matter has correct saturation point and the
    liquid-gas phase transition.
    The quark--gluon phase is described by the bag model with lowest-order perturbative
    corrections. The phase transition boundaries and characteristics of the mixed phase (MP) are found
    by using the Gibbs conditions with the strangeness neutrality constraint.

    To probe sensitivity to the EoS, we have also performed calculations with the EoS of ideal hadron gas
    (EoS--HG). In this case we disregard the excluded volume effects assuming that \mbox{$v_e=0$}.
    To provide stability of initial nuclei we also introduce the Skyrme-like mean field. The parameters of
    EoS--PT and EoS--HG are given in Ref.~\cite{Sat09}.
    Our analysis shows that although generally the EoS with the phase transition is softer than
    the EoS of pure hadron gas, in some regions of thermodynamic parameters the EoS-PT has sound
    velocities of the same order or even larger than those in the EoS--HG.

    Below we consider collisions of gold nuclei. The hydrodynamic simulation is started at the stage
    when two cold Lorentz-contracted nuclei approach each other.
    The initial baryon and energy density profiles of each nucleus
    in its rest frame are described by the Woods-Saxon distribution.
    Due to the stabilizing effect of the Skyrme-like mean field, the initial nuclei may stay
    in equilibrium with vacuum at $P=0$\,. They propagate without distortion until their density
    distributions essentially overlap.

    Below the beam axis is denoted by $z$ and the $x$--axis is chosen along
    the impact parameter vector $\bm{b}$.
    In our simulations we use linear interpolation of tables $P(n,\varepsilon)$
    prepared with fixed steps in $\varepsilon$ and $n$.
    Analogous tables for temperature~($T$), baryon~($\mu$) and strange~($\mu_S$)
    chemical potentials are used for calculating hadronic spectra.
    The numerical solution of fluid--dynamical equations is obtained by using the
    relativistic version of the flux-corrected transport algorithm~\cite{Bor73}.

    To calculate hadronic momentum distributions we apply the approximation
    of instantaneous freeze-out: it is assumed that a sudden transition
    from the local equilibrium to collisionless propagation of particles
    takes place at some isochronous hypersurface (\mbox{$t=t_{\rm fr}={\rm const}$}).
    Within this approximation one can use the following formula~\cite{Coo74}
    for the invariant momentum distribution of the hadronic species $i$
    \bel{spec}
    E\hsp\frac{d^{\hsp 3}N_i}{\hspace*{-4pt}d^{\hsp 3}p}=\frac{d^{\hsp 3}N_i}{dy\hsp d^{\hsp 2}p_{\hsp T}}=
    \frac{g_iE}{(2\pi)^{\hsp 3}}
    \int dV\left\{\exp\left(\frac{p_\nu u^\nu-\mu_i}
    {T}\right)\pm 1\right\}^{-1},
    \ee
    where \mbox{$p^{\hsp\mu}=(E,\bm{p})^{\mu}$} is the 4--momentum of the particle,
    $y$ and $\bm{p}_{\hsp T}$ are, respectively, its longitudinal rapidity and
    transverse momentum, $g_i$ is the statistical weight of i-th hadrons.
    Plus and minus in \re{spec} correspond, respectively, to fermions and bosons.
    Using conditions of chemical equilibrium one can express the particle's
    chemical potential~$\mu_i$ through the baryon and strange
    chemical potentials as follows
    \bel{cceq}
    \mu_i=B_i\mu+S_i\mu_S\,,
    \ee
    where $B_i$ and $S_i$ are, respectively, the baryon and strangeness number
    of species $i$\,.

    In addition to contributions of ''thermal'' nucleons and pions, which are calculated
    directly by using Eqs.~(4)--(5), we also take into account resonance decays, e.g.
    \mbox{$\Delta\to N\pi$}, \mbox{$\rho\to 2\pi$},\ldots
    We use the standard formulae for two-body decays of resonances in the zero width approximation
    (the decays into more than two hadrons are treated approximately~\cite{Sat07}).
    All known resonances with masses up to 2 GeV are taken into account.

    The transverse collective flows of matter created in heavy--ion collisions,
    are rather sensitive to the EoS. Especially useful is the elliptic flow
    characterized by the parameter $v_2$~\cite{Oll92}.
    To discuss qualitatively possible differences between
    the EoS--PT and EoS--HG, below we calculate the so--called momentum anisotropy parameter $\epsilon_p$
    which characterizes the flow asymmetry in the azimuthal plane. Following Ref.~\cite{Boz09}
    we define this quantity as
    \bel{anism}
    \epsilon_p=\frac{\ds\int dx\hsp dy\left(T^{\hsp xx}-T^{\hsp yy}\right)}
    {\ds\int dx\hsp dy\left(T^{\hsp xx}+T^{\hsp yy}\right)}\,,
    \ee
    where $T^{\hsp xx}, T^{\hsp yy}$ are the components of the energy-momentum tensor
    in the azimuthal plane \mbox{$z=0$}.
    The approximate relation \mbox{$\epsilon_p\simeq 2\hsp v_2$} has been obtained in Ref.~\cite{Kol00}
    from (2+1)--hydrodynamical simulations of Pb+Pb collisions at SPS and RHIC energies.

\section{Results}
    First, let us discuss thermodynamic characteristics of matter created in relativistic Au+Au collisions.
    Our goal is to find differences between the results obtained with \mbox{EoS--HG} and EoS--PT.
    Figures 1--2 represent dynamical
    trajectories of matter in the central box \mbox{({$|x|,\,|y|,\,|z|/\gamma_0\,<\,1\,{\rm fm}$})}
    produced in central (b=0) Au+Au collisions at
    different bombarding energies $\ela$.
    Shown are the results for $n-\varepsilon$ and $\mu-T$ planes.
    At \mbox{$\ela\gtrsim 5$ AGeV} the calculation with EoS--PT predicts
    larger values of
    $\varepsilon$ and $n$ as compared to EoS--HG.
    In this case one can also see longer life-times of states
    with maximal compression and delay in transition to expansion stages.
    Note that calculations with EoS--PT predict a zigzag-like behavior of the
	trajectories in the $\mu-T$ plane
    with slightly raising temperature in the MP as a function of time.
	It is important to note that the final states of hadronic matter in the central cell
	are very similar for two EoS used in the calculations. In other words, all
	differences in the early dynamics are practically washed out in the final stage.

    According to our analysis, especially interesting is the region of
	bombarding energies around \mbox{$\ela\sim 10$ AGeV}.
    At such energies we find an enhanced sensitivity of collective flows
    and particle spectra to the phase transition.
    Figure~\ref{fig3} represents the time evolution of the momentum anisotropy $\epsilon_p$
    in semicentral Au+Au collisions at different $\ela$\,. One can see that
    the momentum anisotropy is rather sensitive to
    EoS at $\ela\lesssim 20$ AGeV. The asymptotic values of $\epsilon_p$
    at $\ela\gtrsim 10$ AGeV are larger in calculations with the deconfinement phase transition.
    This agrees with the similar conclusion of Ref.~\cite{Pet10}.

    To estimate sensitivity of these results to the choice of the freeze-out time, we
    determine the time moment when the energy density
    in a central box becomes smaller than a certain freeze-out value $\varepsilon_{\rm fr}$.
    In Fig.~\ref{fig3}
    we mark the values of $\epsilon_p$ corresponding to different values
    of $\varepsilon_{\rm fr}$. Figure 4 shows our results for the excitation function $\epsilon_p (\ela)$.
    The lines connect the values~$\epsilon_p$ taken at
    freeze-out times corresponding to \mbox{$\varepsilon_{\rm fr}=0.4$ GeV/fm$^3$}.
    In the case of EoS--PT we predict a non-monotonic dependence of
    $\epsilon_p (\ela)$ with maximum at \mbox{$\ela\simeq 10$ AGeV}. On the other hand, the existing
    experimental data on the elliptic flow in Au+Au and Pb+Pb collisions at AGS and low SPS energies
    do not show a non-monotonic behavior.
    This may imply that elliptic flows are strongly suppressed by viscosity effects in the considered
    energy domain.

    In Figs. 5--6 we present our results for proton and pion rapidity distributions
    in central Au+Au collisions at $\ela\sim 10$ A GeV.
    The proton and $\pi^-$  distributions
    are obtained from nucleon and pion spectra by introducing the additional factors 1/2 and
    1/3 respectively. The hadronic spectra were calculated by using Eqs. (4)--(5).
    The parameter  $t_{\rm fr}$ is chosen
    to achieve the best fit of experimental data.
    To take into account the mean-field
    potential, we shift the chemical potentials of baryons by $-U(n)$.
    In the case of EoS--PT we also introduce the excluded volume corrections. This is done
    by replacing $\mu_i \rightarrow \mu_i - v_{e}P_K$, where $P_K$ is the kinetic part of pressure~\cite{Sat09}.
    According to our calculations, these corrections are more important for pion spectra.

    Figure~\ref{fig5} shows the rapidity distributions
    of protons in central Au+Au collision at \mbox{$\ela=10.7$ AGeV}.
    The calculation with the phase transition predicts
    noticeably broader rapidity distributions.
    In this case the agreement with experimental data is better as compared
    with the EoS--HG.

    The $\pi^-$ rapidity distribution for the same reaction is shown in Fig.~\ref{fig6}.
    Again, calculations with the EoS--PT give broader distributions than those with the EoS--HG,
    however this difference is not so large as for protons.
    In the case of EoS-PT the observed pion spectra may be reproduced only if one
    takes a smaller freeze-out time as compared to protons. We expect that this differences
	would decrease if we would take smaller values of the excluded volume for pions.

\section{Conclusions}
    We have studied the sensitivity of the elliptic flow and particle spectra to the deconfinement
    phase transition in heavy-ion collisions.
    Our analysis shows that at FAIR energies maximal values of energy- and baryon densities
    in the central box of the colliding system are significantly larger
    if the QGP is formed at some intermediate stage of a heavy-ion collision.
    It it shown that the collective flow parameters are especially sensitive to the
	EoS around $\ela \simeq 10$ AGeV.
    At such energies the calculations with the deconfinement phase transition predict enhanced
    elliptic flows and broader proton rapidity distributions as compared with the purely hadronic scenario.
    
\begin{acknowledgments}
    The authors thank D.H.~Risch\-ke for providing us the one--fluid 3D code,
    M.I.~Gorenstein, P.~Huovinen, Yu.B.~Ivanov, H.~Niemi, D.Yu.~Peressounko,
    and V.D.~Toneev for useful discussions. The computational resources were
    provided by the Center for Scientific Computing (Frankfurt am Main) and by
    the Grid Computing Center (the Kurchatov Institute, Moscow).
    This work was supported in part by the DFG \mbox{grant 436 RUS 113/957/0--1},
    the Helmholtz International Center for FAIR~(Germany) and
    the grants RFBR 09--02--91331 and NSH--7235.2010.2 (Russia).
\end{acknowledgments}

%\newpage

\newpage
        % Fig1
        \begin{figure*}[htb!]
        \vspace*{-5cm}
        \hspace*{-0.5cm}\includegraphics[width=0.8\textwidth]{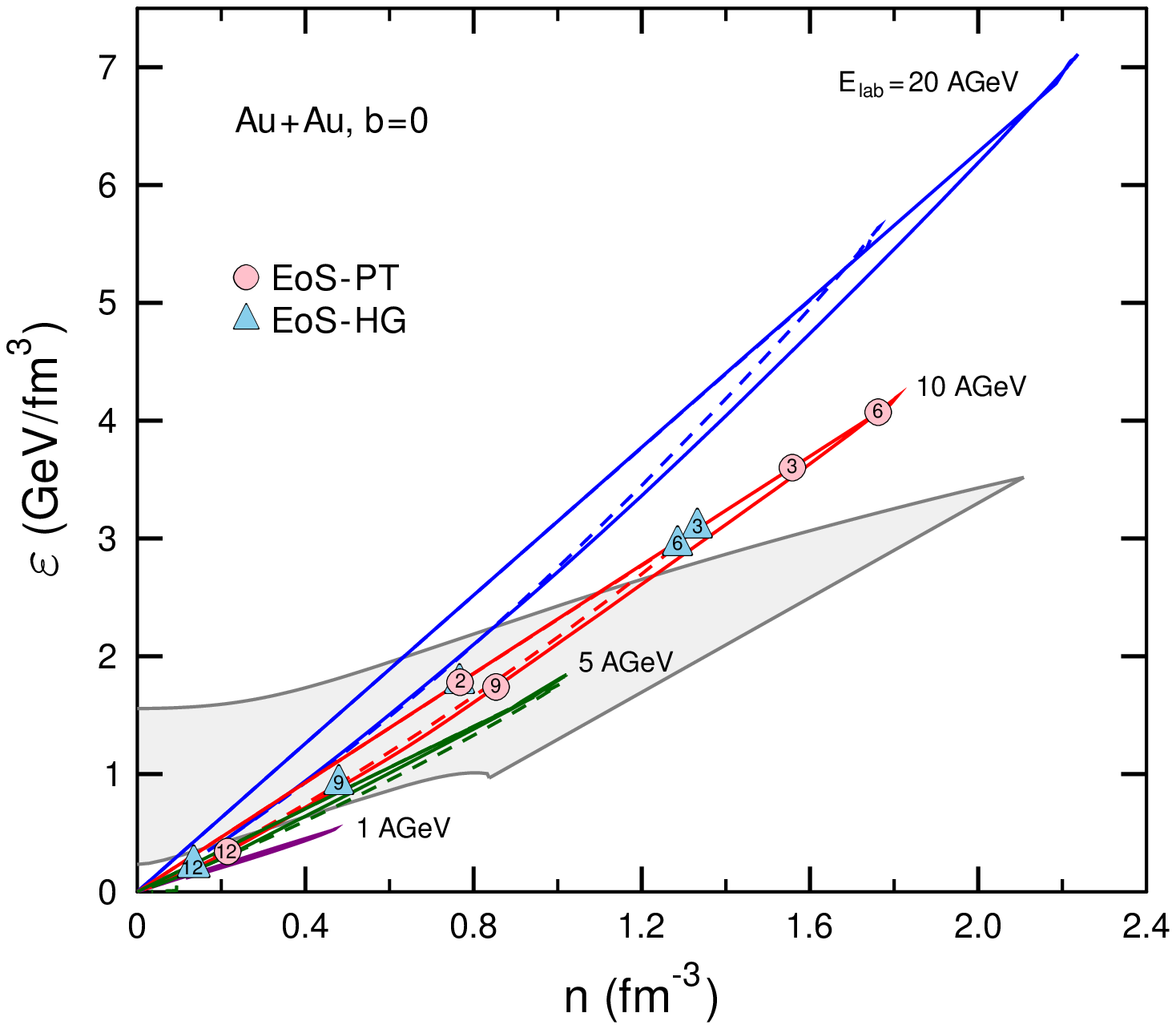}
        \vspace*{-5.5cm}
        \caption{
        Time evolution of matter in central Au+Au collision at different bombarding energies~$\ela$.
        Shown are values of energy and baryon densities averaged over the central box.
        Dashed and solid lines correspond to EoS--HG and EoS--PT, respectively. Numbers
        in circles and triangles give the c.m. time in fm/$c$.
        Shading shows the MP region of deconfinement phase transition.
        }
        \label{fig1}
        \end{figure*}

        % Fig2
        \begin{figure*}[htb!]
        \vspace*{-7cm}
        \centerline{\includegraphics[width=0.90\textwidth]{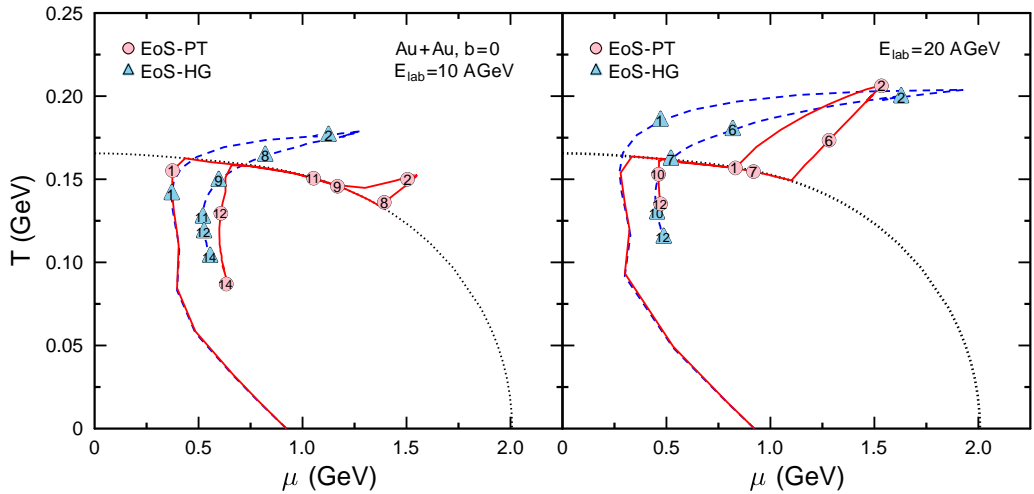}} %0.95
        \vspace*{-7.5cm}
        \caption{
        Time evolution of matter (central cell) in the \mbox{$\mu-T$} plane.
        Left and right panels correspond to central Au+Au collisions
        at $\ela=10$ and 20 AGeV, respectively.
        Dashed (solid) lines show the results for EoS--HG (EoS--PT).
        Dotted lines show the MP region. Numbers
        in circles and triangles give time values in fm/$c$.
        }
        \label{fig2}
        \end{figure*}

\newpage
        % Fig3
        \begin{figure*}[htb!]
        \vspace*{-2.5cm}
        \centerline{\includegraphics[width=1.2\textwidth]{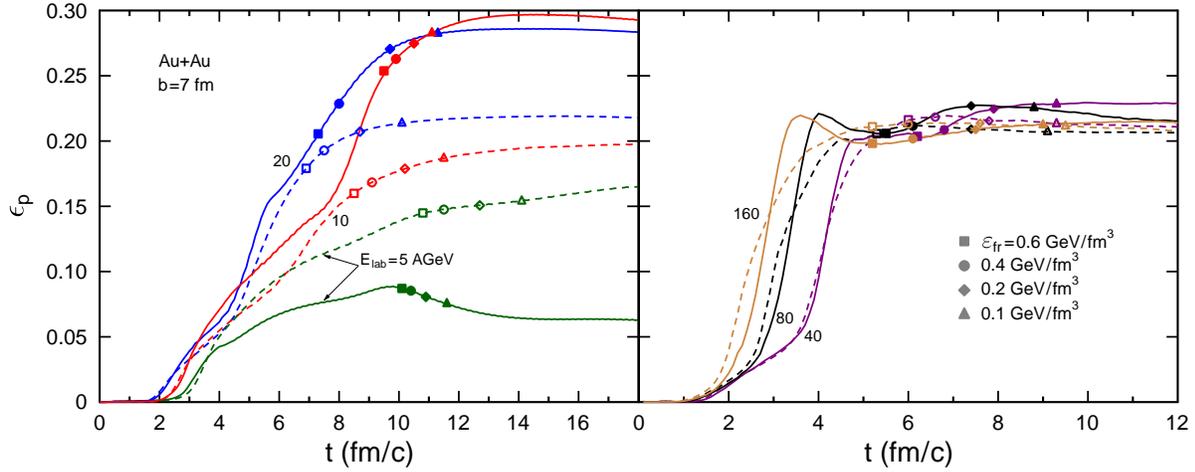}}
        \vspace*{-3cm}
        \caption{
        Time dependence of the momentum anisotropy
        in Au+Au collisions for different bombarding energies ($b=7$ fm).
        The solid (dashed) lines correspond to the EoS--PT (EoS--HG). Markers
        show time moments when energy density in the central box
        becomes lower than certain values $\varepsilon_{\rm{fr}}$ indicated in the the right panel.
        }
        \label{fig3}
        \end{figure*}

        % Fig4
        \begin{figure*}[htb!]
        \vspace*{-4.2cm}
        \centerline{\includegraphics[width=0.65\textwidth]{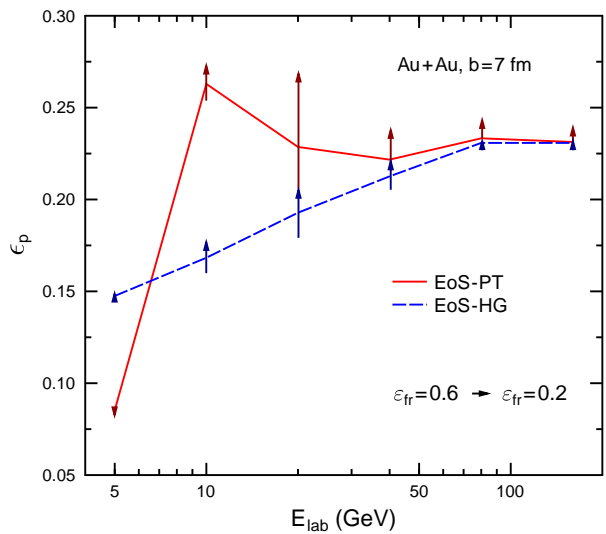}}
        \vspace*{-4cm}
        \caption{
        Excitation function of momentum anisotropy in Au+Au collisions with
        \mbox{$b=7$ fm}. The~solid (dashed) lines correspond to the EoS--PT (EoS--HG).
        Arrows show possible shifts of $\epsilon_p$-values for different choices
        of freeze-out energy density $\varepsilon_{\rm fr}$ between 0.2 and 0.6 GeV/fm$^3$.
        }
        \label{fig4}
        \end{figure*}

\newpage
        % Fig5
        \begin{figure*}[htb!]
        \vspace*{-4.2cm}
        \centerline{\includegraphics[width=0.70\textwidth]{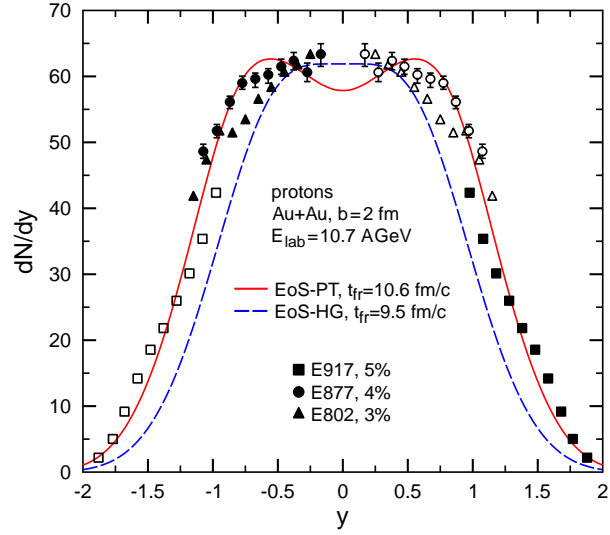}}
        \vspace*{-4cm}
        \caption{
        Proton rapidity distributions for central Au+Au collision at $\ela = 10.7$ AGeV.
        Full~symbols are experimental data~\cite{Ahl98,Bac01,Bar00}.
        Open symbols are obtained by reflection with respect to midrapidity.
        }
        \label{fig5}
        \end{figure*}

         % Fig6
        \begin{figure*}[htb!]
        \vspace*{-4.2cm}
        \centerline{\includegraphics[width=0.70\textwidth]{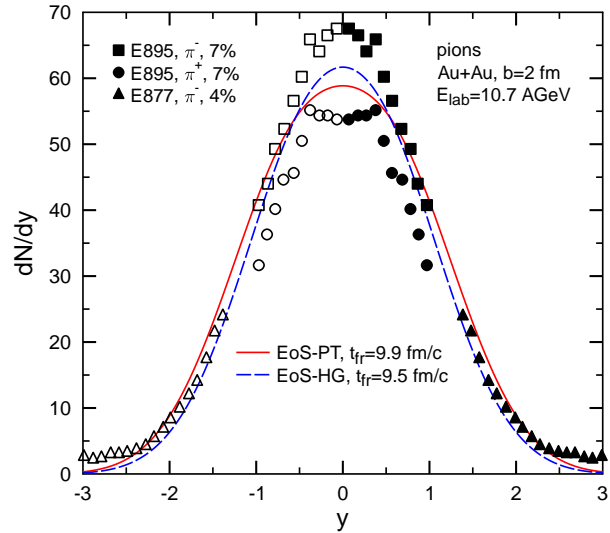}}
        \vspace*{-4cm}
        \caption{
        Same as Fig.~\ref{fig5} but for rapidity distributions of $\pi^-$ mesons.
        Experimental data are taken from Refs.~\cite{Bar00,Kla03}.
        }
        \label{fig6}
        \end{figure*}

\end{document}